\documentclass[12pt,thmsa]{article}
\usepackage{amsfonts}
\usepackage{amssymb}
\usepackage{amsmath}
\usepackage{sw20aip}

\input tcilatex
\begin{document}

\title{Comment on \textquotedblleft The spin symmetry for deformed
generalized P\"{o}schl-Teller potential\textquotedblright }
\author{L. Aggoun, F. Benamira and L. Guechi \\
Laboratoire de Physique Th\'{e}orique \and D\'{e}partement de Physique, \and %
Facult\'{e} des Sciences Exactes, Universit\'{e} Mentouri, \and Route d'Ain
El Bey, Constantine, Algeria.}
\maketitle

\begin{abstract}
In this comment, we show that the solutions of the s-wave Dirac equation for
deformed generalized P\"{o}schl-Teller potential obtained by Wei et al are
valid only for $q\geq 1$ and $\frac{1}{2\alpha }\ln q<r<\infty $.
Consequently, numerical results contained in Table 1 are entirely wrong
because the energy equation (20) in Ref.\cite{Dong} is not good for $q=0,3$.
\ When $0<q<1$, we prove that the energy eigenvalues for the bound states
are given by the solution of a transcendental equation involving the
hypergeometric function. To test our results, the Morse potential is
considered as limiting case.


\end{abstract}

In a recent paper\cite{Dong} , the s-wave Dirac equation is solved for a
deformed generalized P\"{o}schl-Teller potential. The authors of Ref.\cite%
{Dong} , start with the following Dirac equation for a nucleon with mass $M$
moving in a scalar potential $S(r)$ and a vector potential $V(r)$:

\begin{equation}
\left\{ \widehat{\mathbf{\alpha }}\widehat{\mathbf{p}}+\widehat{\beta }\left[
M+S(r)\right] \right\} \psi (\mathbf{r})=\left[ E-V(r)\right] \psi (\mathbf{r%
}),  \label{WD1}
\end{equation}%
and define the Dirac spinors as 
\begin{equation}
\psi _{n_{r}k}(\mathbf{r})=\frac{1}{r}\left( 
\begin{array}{c}
F_{n_{r}k}(r)Y_{jm}^{l}(\theta ,\phi ) \\ 
iG_{n_{r}k}(r)Y_{jm}^{\widetilde{l}}(\theta ,\phi )%
\end{array}%
\right) ,  \label{WD2}
\end{equation}%
obtained from Eq. (\ref{WD1}), with $F_{n_{r}k}(r)$ and $G_{n_{r}k}(r)$ are
the upper and lower component radial wave functions, respectively. For the
s-wave ($k=-1$) case, the upper and lower components are determined by the
system of differential equations%
\begin{equation}
\left( \frac{d}{dr}-\frac{1}{r}\right) F_{n_{r},-1}(r)=\left(
M+E_{n_{r},-1}-\triangle (r)\right) G_{n_{r},-1}(r),  \label{WD3a}
\end{equation}%
\begin{equation}
\left( \frac{d}{dr}+\frac{1}{r}\right) G_{n_{r},-1}(r)=\left(
M-E_{n_{r},-1}+\Sigma (r)\right) F_{n_{r},-1}(r),  \label{WD3b}
\end{equation}%
where $\triangle (r)=$ $V(r)-S(r)$ and $\Sigma (r)=V(r)+S(r).$ Under the
condition of exact spin symmetry, i.e, $\frac{d\triangle (r)}{dr}=0$ or $%
\triangle (r)=C=$Constant, they obtain the following equation for the upper
component%
\begin{equation}
H_{n_{r},-1}F_{n_{r},-1}(r)=\widetilde{E}_{n_{r},-1}F_{n_{r},-1}(r).
\label{WD4}
\end{equation}%
This equation is formally the same as the Schr\"{o}dinger equation in which%
\begin{equation}
H_{n_{r},-1}=-\frac{d^{2}}{dr^{2}}+\left( M+E_{n_{r},-1}-C\right) \Sigma (r)
\label{WD5}
\end{equation}%
plays the role of the Hamiltonian and its eigenvalue is%
\begin{equation}
\widetilde{E}_{n_{r},-1}=E_{n_{r},-1}^{2}-M^{2}+C\left(
M-E_{n_{r},-1}\right) .  \label{WD6}
\end{equation}%
The sum potential $\Sigma (r)$ is taken as the deformed generalized P\"{o}%
schl-Teller potential,

\begin{equation}
V_{q}(r)=\frac{V_{1}-V_{2}\cosh _{q}(\alpha r)}{\sinh _{q}^{2}(\alpha r)},
\label{WD7}
\end{equation}%
where $V_{1}$ and $V_{2}$ are positive constants such that $V_{1}>V_{2}$ and 
$q$ is a deformation parameter which can take any real positive value.

In this comment, we would like to point out that there is a serious point
which invalidates partially Eqs.(20) and (21) in Ref.\cite{Dong} , and which
has to do with the definition of the potential (\ref{WD7}). Note that this
potential has a strong singularity at the point $r=r_{0}=\frac{1}{2\alpha }%
\ln q$, for $q\geq 1$, which creates an impenetrable barrier. Then, if we
assume that $q$ is positive as in Ref.\cite{Dong} , the treatment of Eq.(\ref%
{WD4}) should be re-examined. Three cases must be distinguished:

(i) $q\geq 1.$

In this case, the particle motion is forced to take place on the half-line $%
r>r_{0}.$ The solution (21) in Ref.\cite{Dong} of Eq.(\ref{WD4}) with the
energy equation (20) in Ref.\cite{Dong} is physically acceptable as upper
component of radial wave functions \ since it fulfills, at the same time,
the boundary conditions

\begin{equation}
F_{n_{r},-1}^{q\geq 1}(r_{0})=0,  \label{WD8}
\end{equation}%
and

\begin{equation}
F_{n_{r},-1}^{q\geq 1}(r)\underset{r\rightarrow \infty }{\rightarrow }0,
\label{WD9}
\end{equation}%
contrary to the statement made in Ref.\cite{Dong} . It is important to note
that the conditions (\ref{WD8}) , (\ref{WD9}) and their likes for $%
G_{n_{r},-1}^{q\geq 1}(r)$ ensure the hermiticity of the effective
Hamiltonian given by Eq.(\ref{WD5}). Nevertheless, we prefer deriving
directly the solution of Eq.(\ref{WD4}) \ in the form of orthogonal
polynomials. With the new variable

\begin{equation}
z=\tanh _{q^{\frac{1}{2}}}^{2}\left( \frac{\alpha }{2}r\right) ,
\label{WD10}
\end{equation}%
and the new function $\varphi _{n_{r},-1}\left( z\right) $ defined by the
relation

\begin{equation}
F_{n_{r},-1}^{q\geq 1}(r)=z^{\lambda }\left( 1-z\right) ^{\eta }\varphi
_{n_{r},-1}\left( z\right) ,  \label{WD11}
\end{equation}%
where $\lambda ,\eta >0$, the Schr\"{o}dinger equation\ (\ref{WD4}) is
reduced to the Gauss's hypergeometric differential equation\cite{Gradshtein}

\begin{equation}
\!\!\!\!\!\!\!\!\left[ z\left( 1-z\right) \frac{d^{2}}{dz^{2}}\!\!+\!\!\left[
\!\!\quad \!\!\!2\lambda \!\!+\!\!\frac{1}{2}\!-\!\!\left( 2\lambda +2\eta
+\!\!\frac{1}{2}\right) z\!\!\,\right] \frac{d}{dz}\!\!-\!\!\left( \!\lambda
\!+\!\eta \!+\!\frac{1}{4}\!\right) ^{2}\!+\!\frac{1}{4\alpha ^{2}}\left( 
\frac{\widetilde{V}_{1}}{q}\!+\!\frac{\widetilde{V}_{2}}{\sqrt{q}}\right)
\!\!+\!\frac{1}{16}\!\!\,\right] \varphi _{n_{r},-1}(z)=0.  \label{WD12}
\end{equation}%
Since $\lambda $ and $\eta $ are positive, the solution of this equation,
for which the boundary condition (\ref{WD8}) and (\ref{WD9}) are fulfilled,
can be written%
\begin{equation}
\varphi _{n_{r},-1}\left( z\right) =A\text{ }_{2}F_{1}(-n_{r},n_{r}+2\lambda
+2\eta +\frac{1}{2},2\lambda +\frac{1}{2};z),  \label{WD14}
\end{equation}%
and $A$ is a constant factor. Thus, the upper component of the radial wave
functions is 
\begin{eqnarray}
F_{n_{r},-1}^{q\geq 1}(r) &=&A\left( \frac{q^{\frac{1}{4}}}{\cosh _{q^{\frac{%
1}{2}}}\left( \frac{\alpha r}{2}\right) }\right) ^{2\eta }\tanh _{q^{\frac{1%
}{2}}}^{2\lambda }\left( \frac{\alpha r}{2}\right) \text{ }  \notag \\
&&\times _{2}F_{1}\left( -n_{r},n_{r}+2\lambda +2\eta +\frac{1}{2},2\lambda +%
\frac{1}{2};\tanh _{q^{\frac{1}{2}}}^{2}\left( \frac{\alpha r}{2}\right)
\right) ,  \label{WD15}
\end{eqnarray}%
where

\begin{equation}
\left\{ 
\begin{array}{c}
\lambda +\eta +\frac{1}{4}\left( 1-\sqrt{1+\frac{4}{\alpha ^{2}}\left( \frac{%
\widetilde{V}_{1}}{q}+\frac{\widetilde{V}_{2}}{\sqrt{q}}\right) }\right) =-%
\text{\ }n_{r}, \\ 
\lambda =\frac{1}{4}\left( 1+\sqrt{1+\frac{4}{\alpha ^{2}}\left( \frac{%
\widetilde{V}_{1}}{q}-\frac{\widetilde{V}_{2}}{\sqrt{q}}\right) }\right) ,%
\text{ \ \ }\eta =\frac{1}{\alpha }\sqrt{-\widetilde{E}_{n_{r},-1}}, \\ 
\text{ }\widetilde{V}_{1}=\left( M+E_{n,-1}-C\right) V_{1},\text{ \ \ \ }%
\widetilde{V}_{2}=\left( M+E_{n,-1}-C\right) V_{2}.%
\end{array}%
\right.  \label{WD16}
\end{equation}%
Using the link between the Jacobi polynomials and the hypergeometric
function (see Ref.\cite{Gradshtein} , formula (8.962.1), p. 1036)%
\begin{equation}
P_{n}^{(\alpha ,\beta )}(x)=\frac{\Gamma (n+\alpha +1)}{n!\Gamma (\alpha +1)}%
\text{ }_{2}F_{1}\left( -n,n+\alpha +\beta +1,\alpha +1;\frac{1-x}{2}\right)
,  \label{WD17}
\end{equation}%
we can also express (\ref{WD15}) in the form%
\begin{equation}
F_{n_{r},-1}^{q\geq 1}(r)=N_{n_{r},-1}\left( \frac{q^{\frac{1}{4}}}{\cosh
_{q^{\frac{1}{2}}}\left( \frac{\alpha r}{2}\right) }\right) ^{2\eta }\tanh
_{q^{\frac{1}{2}}}^{2\lambda }\left( \frac{\alpha r}{2}\right)
P_{n}^{(2\lambda -\frac{1}{2},2\eta )}\left( 1-2\tanh _{q^{\frac{1}{2}%
}}^{2}\left( \frac{\alpha r}{2}\right) \right) ,  \label{WD18}
\end{equation}%
where the constant $N_{n_{r},-1}$ is determined from the normalisation
condition 
\begin{equation}
\int_{r_{0}}^{\infty }\left[ \left( F_{n_{r},-1}^{q\geq 1}(r)\right)
^{2}+\left( G_{n_{r},-1}^{q\geq 1}(r)\right) ^{2}\right] dr=1.  \label{WD19}
\end{equation}

(ii) $0<q<1.$

The potential (\ref{WD7}) \ is continuous on the whole interval $%
\mathbb{R}
^{+}$. In order to solve \ (\ref{WD4}), it is convenient to transform the
radial variable $r\in \left( 0,\infty \right) $ into a variable $z\in \left(
0,4q^{\frac{1}{2}}/(1+q^{\frac{1}{2}})^{2}\right) $ by%
\begin{equation}
z=\frac{q^{\frac{1}{2}}}{\cosh _{q^{\frac{1}{2}}}^{2}\left( \frac{\alpha }{2}%
r\right) },  \label{WD20}
\end{equation}%
and we look for a solution in the form 
\begin{equation}
F_{n_{r},-1}^{0<q<1}(r)=z^{\eta }\left( 1-z\right) ^{\lambda }\varphi
_{n_{r},-1}\left( z\right) ,  \label{WD21}
\end{equation}%
in which, on account of the boundary conditions

\begin{equation}
F_{n_{r},-1}^{0<q<1}(0)=0,  \label{WD22}
\end{equation}%
and%
\begin{equation}
\underset{r\rightarrow \infty }{\lim }F_{n_{r},-1}^{0<q<1}(r)=0,
\label{WD23}
\end{equation}%
$\eta $ and $\lambda $ are similarly given by Eqs. (\ref{WD16}). As in the
previous case, the boundary conditions (\ref{WD22}) and (\ref{WD23}) and
their likes for $G_{n_{r},-1}^{0<q<1}(r)$ come into existence by requiring
that the effective Hamiltonian given by (\ref{WD6}) is hermitian. Proceeding
in the same way as for the case where $q\geq 1$, the upper component $%
F_{n_{r},-1}^{0<q<1}(r)$ satisfying Eqs. (\ref{WD4}), (\ref{WD22}) and (\ref%
{WD23}) is thus given by (\ref{WD21}) as 
\begin{eqnarray}
F_{n_{r},-1}^{0<q<1}(r) &=&N\left( \frac{q^{\frac{1}{2}}}{\cosh _{q^{\frac{1%
}{2}}}^{2}\left( \frac{\alpha }{2}r\right) }\right) ^{\eta }\left( 1-\frac{%
q^{\frac{1}{2}}}{\cosh _{q^{\frac{1}{2}}}^{2}\left( \frac{\alpha }{2}%
r\right) }\right) ^{\lambda }  \notag \\
&&\times _{2}F_{1}\left( a,b,c;\frac{q^{\frac{1}{2}}}{\cosh _{q^{\frac{1}{2}%
}}^{2}\left( \frac{\alpha }{2}r\right) }\right) ,  \label{WD24}
\end{eqnarray}%
\ where

\begin{equation}
\left\{ 
\begin{array}{c}
a=\eta +\lambda +\frac{1}{4}\left( 1-\sqrt{1+\frac{4}{\alpha ^{2}q}\left( 
\widetilde{V}_{1}+\widetilde{V}_{2}\sqrt{q}\right) }\right) , \\ 
b=\eta +\lambda +\frac{1}{4}\left( 1+\sqrt{1+\frac{4}{\alpha ^{2}q}\left( 
\widetilde{V}_{1}+\widetilde{V}_{2}\sqrt{q}\right) }\right) , \\ 
c=2\eta +1\text{ },\ \ \text{\ \ }%
\end{array}%
\right.  \label{WD25}
\end{equation}%
and $N$ is a constant factor. From the boundary condition (\ref{WD22}), we
obtain the transcendental quantization equation for the bound state energy
levels $E_{n_{r},-1}$:

\begin{equation}
_{2}F_{1}\left( a,b,c;\frac{4q^{\frac{1}{2}}}{\left( 1+q^{\frac{1}{2}%
}\right) ^{2}}\right) =0,  \label{WD26}
\end{equation}%
which can be solved numerically. Consequently, the values of $E_{n_{r},-1}$
calculated from Eq.(20) in Ref.\cite{Dong} for $q=0.3$ and presented in
table $1$ of Ref.\cite{Dong} \ are wrong. The correct numerical values of $%
E_{n_{r},-1}$ for $0<q<1$ have to be determined from Eq.(\ref{WD26}).

(iii) The radial Morse potential

If $q$ tends to zero, the deformed generalized P\"{o}schl-Teller potential (%
\ref{WD7}) tends to

\begin{equation}
V_{0}(r)=4V_{1}e^{-2\alpha r}-2V_{2}e^{-\alpha r},  \label{WD27}
\end{equation}%
which is the so-called radial Morse potential with the parameters $V_{1}$
and $V_{2}$ defined by $V_{1}=\frac{D_{e}}{4}e^{2\alpha r_{e}}$ and $%
V_{2}=D_{e}e^{\alpha r_{e}}$, where $D_{e}$ is the depth of the potential
well and $r_{e}$ is the equilibrium distance of the two nuclei.

In this case, it can be seen from Eqs.(\ref{WD16}) and (\ref{WD25}) that

\begin{equation}
\left\{ 
\begin{array}{c}
\lambda \underset{q\rightarrow 0}{\simeq }\frac{1}{4}\left( 1-\frac{%
\widetilde{V}_{2}}{\alpha \sqrt{\widetilde{V}_{1}}}+\frac{2\sqrt{\widetilde{V%
}_{1}}}{\alpha \sqrt{q}}\right) \underset{q\rightarrow 0}{\rightarrow }%
\infty , \\ 
a\underset{q\rightarrow 0}{\simeq }\frac{1}{2}+\eta -\frac{\widetilde{V}_{2}%
}{2\alpha \sqrt{\widetilde{V}_{1}}}, \\ 
b\underset{q\rightarrow 0}{\simeq }\frac{1}{2}+\eta +\frac{\sqrt{\widetilde{V%
}_{1}}}{\alpha \sqrt{q}}\underset{q\rightarrow 0}{\simeq }\frac{\sqrt{%
\widetilde{V}_{1}}}{\alpha \sqrt{q}}\underset{q\rightarrow 0}{\rightarrow }%
\infty .%
\end{array}%
\right.  \label{WD28}
\end{equation}%
On the other hand, by using the formula\cite{Landau}%
\begin{equation}
\underset{\beta \rightarrow \infty }{\lim }\text{ }_{2}F_{1}\left( \alpha
,\beta ,\gamma ;\frac{z}{\beta }\right) =\text{ }_{1}F_{1}(\alpha ,\gamma
;z),  \label{WD29}
\end{equation}%
it is easy to show that, in the limiting case $q\rightarrow 0$, the upper
component of the wave functions (\ref{WD24}) becomes%
\begin{eqnarray}
F_{n_{r},-1}^{0<q<1}(r)\underset{q\rightarrow 0}{\simeq }F_{n_{r},-1}(r)
&=&C\exp \left( -\sqrt{-\widetilde{E}_{n_{r},-1}}r\right) \exp \left( -\frac{%
4\sqrt{\widetilde{V}_{1}}}{\alpha }e^{-\alpha r}\right)  \notag \\
&&\times _{1}F_{1}\left( \frac{1}{2}-\frac{\widetilde{V}_{2}}{2\alpha \sqrt{%
\widetilde{V}_{1}}}+\frac{1}{\alpha }\sqrt{-\widetilde{E}_{n_{r},-1}},\frac{2%
}{\alpha }\sqrt{-\widetilde{E}_{n_{r},-1}}+1;\frac{4\widetilde{V}_{1}}{%
\alpha }e^{-\alpha r}\right) ,  \notag \\
&&  \label{WD30}
\end{eqnarray}%
and the transcendental quantization condition for the bound state energy
levels (\ref{WD26}) takes the form%
\begin{eqnarray}
_{2}F_{1}\left( a,b,c;\frac{4q^{\frac{1}{2}}}{\left( 1+q^{\frac{1}{2}%
}\right) ^{2}}\right) \underset{q\rightarrow 0}{\simeq }\text{ }%
_{1}F_{1}\left( \frac{1}{2}-\frac{\widetilde{V}_{2}}{2\alpha \sqrt{%
\widetilde{V}_{1}}}+\frac{1}{\alpha }\sqrt{-\widetilde{E}_{n_{r},-1}},\frac{2%
}{\alpha }\sqrt{-\widetilde{E}_{n_{r},-1}}+1;\frac{4\widetilde{V}_{1}}{%
\alpha }\right) &=&0.  \notag \\
&&  \label{WD31}
\end{eqnarray}%
As $\widetilde{V}_{1}\rightarrow \infty ,$ from the asymptotic behaviour of
the confluent hypergeometric function\cite{Flugge} , it follows that 
\begin{equation}
\frac{1}{2}-\frac{\widetilde{V}_{2}}{2\alpha \sqrt{\widetilde{V}_{1}}}+\frac{%
1}{\alpha }\sqrt{-\widetilde{E}_{n_{r},-1}}=-n_{r};\text{ \ }n_{r}=0,1,2,...%
\text{ . \ }  \label{WD32}
\end{equation}%
From this condition, one obtains the energy equation%
\begin{equation}
\widetilde{E}_{n_{r},-1}=-\alpha ^{2}\left( n_{r}-\frac{\widetilde{V}_{2}}{%
2\alpha \sqrt{\widetilde{V}_{1}}}+\frac{1}{2}\right) ^{2}.  \label{WD33}
\end{equation}%
This limiting case constitutes a consistency check for the correctness of
our results.

In conclusion, the numerical energy values given in Table 1 are incorrect
because for $0<q<1$, they are determined from the numerical solution of the
transcendental equation (\ref{WD26}).

\end{document}